\documentclass[prl,twocolumn,showpacs]{revtex4}
\usepackage{amsmath}
\usepackage{amsfonts}

\begin{document}
\title{How significant are the known collision and element distinctness quantum algorithms?}
\author{Lov Grover}
\email{lkgrover@bell-labs.com}
\author{Terry Rudolph}
\email{rudolpht@bell-labs.com}
\affiliation{Bell Labs, 600-700 Mountain Ave., Murray Hill, NJ 07974,
  U.S.A.}

\date{\today}

\begin{abstract}
Quantum search is a technique for searching $N$ possibilities in
only $O(\sqrt{N})$ steps. It has been applied in the design of
quantum algorithms for several structured problems. Many of these
algorithms require significant amount of quantum hardware. In this
paper we observe that if an algorithm requires $O(P)$ hardware, it
should be considered significant if and only if it produces a
speedup of at least $O\left(\sqrt{P}\right) $ over a simple
quantum search algorithm. This is because a speedup of
$O\left(\sqrt{P}\right) $ can be trivially obtained by dividing
the search space into $O(P)$ separate parts and handing the
problem to independent processors that do a quantum search. We
argue that the known algorithms for collision and element
distinctness fail to be non-trivial in this sense.
\end{abstract}

\maketitle

\section{Background}

The quantum search algorithm gave a means of searching $N\;$items
in only $\sqrt{N}$ steps \cite{grover96}. Unlike most computer
science applications, this did not require the problem under
consideration to have any structure that the algorithm could make
use of.

It is easy to see that any classical algorithm, whether
probabilistic or deterministic, would need $O(N)$ oracle queries
for unstructured searching - it had generally been assumed that
$O(N)$ steps would be required by \textit{any} algorithm. However,
quantum mechanical systems can be in multiple states
simultaneously and there is no clearly defined bound on how
rapidly they can search. It was proven through subtle properties
of unitary transformations that any quantum computer would need at
least $O(\sqrt{N})$ queries to search $N$ items \cite{bbbv}.
Subsequently it was shown that the number of queries
required by the algorithm was optimal; it can not be improved even by one %
\cite{zalka}.

The technique behind the algorithm is very general and through the
amplitude amplification principle \cite{amp,BBHT}, the algorithm
has been applied to a number of different \emph{structured}
problems, where it has yielded the best known algorithms. In many
of these settings the algorithm requires additional hardware in
the form of memory registers.

\section{Parallelized Quantum Searching}

If we have to search $N$ items for a target state, the quantum
search algorithm takes $O\left(\sqrt{N}\right) $ operations.
Alternatively, we could divide the $N$ items into $P$ groups of
$\frac{N}{P}$ items each and hand each group to an independent
quantum processor each of which runs an independent quantum search
in $O\left(\sqrt{\frac{N}{P}}\right) $ steps. This division gives
a speedup of $\sqrt{P}$ over the quantum search algorithm by using
$P$ processors. Zalka proved that this was the best possible
speedup for the quantum search algorithm using parallelization.
His proof was for unstructured problems. It leaves open the
possibility for better parallel speedups for structured problems.\
However, as we show in this critique,\emph{\ many well known
algorithms fail to meet this simple benchmark}, i.e. the speedup
they get is $\leq $ $\sqrt{P}.$

\section{Quantum Hardware: Processors \&\ Memory}

In traditional classical computing, there are considerable
differences between the requirements that information processors
and information storers (memory) necessarily satisfy. As such, the
physical realizations of these components can be quite distinct -
e.g. transistors make good processors, oriented magnetic domains
make good memory. Memory is normally ``cheaper''. Consequently, it
is common to treat these components as completely different
resources within classical computer science.

Within quantum computing, however, the distinction between the two types of
component is much more blurred. A\ qubit register that must act as quantum
memory (to hold the output of some computation say) is generally required to
remain coherent with the other systems comprising the quantum computer. In
fact, within the standard quantum computational model they must not merely
remain coherent - they must be capable of dynamically coupling to other
quantum systems within the computer via coherent unitary evolution. A
distinction between ``memory'' and ``computer'' qubits could perhaps be
artificially imposed by dictating that memory qubits can only undergo
controlled-NOT or Toffoli gates - this does not, however, seem pragmatically
justifiable. Most realizations of a quantum computer are more easily capable
of single qubit unitary evolution than these two or three qubit gates, and
thus such memory qubits could be trivially extended to processing qubits.

Thus it seems clear to us that analysis of the space/time
complexity of quantum algorithms is best served by simply treating
all required qubits as available for running any aspect of the
algorithm under consideration.

\section{Element distinctness \& Collision Problems}

\subsection{Problem description}

Two problems that contain some structure, and that therefore could
potentially be solved on a quantum computer better than by
exhaustive searching, are the \emph{collision} and \emph{element
distinctness} problems. We focus here on the simplest versions of
these problems.

In the (two-to-one) collision problem, we are given a (black box) function $%
F(x)$ with a domain of (even) cardinality $N,$ and we are asked to determine
whether $F$ is \emph{one-to-one} or \emph{two-to-one. }That is, we know that
either every item in the domain maps onto a unique point or exactly one pair
of items map onto every point in the range - we need to know which it is%
\emph{.} This is an important problem - it is used quite widely in
cryptography. Also a logarithmic time algorithm for this would solve the
graph isomorphism problem.

The element distinctness problem is similar to the collision problem, except
that now \ we have to determine whether there is \emph{any }pair of inputs $%
x,y$ to the function, such that $F(x)=F(y).$

\subsection{Algorithms \& bounds in terms of query complexity}

\subsubsection{Collision: }

A well known classical algorithm (based on the birthday paradox)
can find a collision in $O(\sqrt{N})$ steps, and in space
$O(\sqrt{N}).$ This is because with a high probability there will
be at least one collision if we examine $O(\sqrt{N})$ random
items. Naive quantum searching requires the same amount of time,
but can reduce the space complexity to a constant factor. To see
this, note that we could search every pair of points in the domain
for a possible collision, there are $^{N}C_{2}$ items to be
searched for $N$ target items where there could be a collision.
Quantum searching would require
$O\left(\sqrt{\frac{^{N}C_{2}}{N}}\right) $ which is
$O\left(\sqrt{N}\right) $ steps. This is the same amount of time
as it would take classically, but the space required has been
reduced to a constant.

The first lower bound for the collision problem was obtained by
Aaronson \cite{scott}, refinements by Shih \cite{shih}, Kutin
\cite{kutin} and Ambainis \cite{ambainis} have shown that there is a $\Omega
(N^{1/3})$ lower bound on the number of queries (calls to $F$) for
any quantum algorithm. This bound matches the algorithm of
\cite{BHT}, which is discussed in detail below.

\subsubsection{Element distinctness:}

Classically, it is possible to check whether or not every item in
the domain maps onto a unique item, by sorting the items according
to their function values and then checking adjacent items. This
sorting and checking would take $O(N)$ steps (up to logarithmic
factors) and $O(N)$ memory. If we were to use the quantum search
algorithm naively, we could search every pair of points in the
domain to check whether or not the function assumed distinct
values, there are $^{N}C_{2}$ items to be searched for a single
target item. Quantum searching would require
$O\left(\sqrt{^{N}C_{2}}\right) $ which is $O\left(N\right) $
steps. This is the same amount of time as it would take
classically, but the space required has been reduced to a
constant.  The algorithm of
\cite{buhrman}, discussed in detail below, achieves a complexity $O(N^{3/4}),$ while
Ambainis has recently discovered an algorithm which needs
$O(N^{2/3})$ steps. Quantum mechanically, the best known lower
bound is $\Omega (N^{2/3})$ on the number of queries any quantum
algorithm must make; this is matched by Ambainis' algorithm.

\bigskip

We see that for both problems the lower bounds on the number of
queries required are matched by known algorithms - this may lead
to the belief that both problems are effectively closed.

\section{Three Algorithms}

We now discuss three well known algorithms for the above problems.
As we will show, these algorithms fail to achieve\ more than a
square-root factor speedup over the number of available processing
qubits. Thus they fail the simple criteria we propose for
determining whether an algorithm makes meaningful use of problem
structure.

\subsection{Collision algorithm}

In 1997, in one of the first significant applications of the
search algorithm, Brassard et al discovered an
$O\left(N^{1/3}\right) $ step algorithm for the collision problem
\cite{BHT}. The algorithm selects $O\left(N^{1/3}\right) $ random items, evaluates $F$  and sorts the outputs in $%
O\left(N^{1/3}\right) $ memory. Then it randomly selects
$O\left(N^{2/3}\right) $ items from the remainder. It may be shown
that with a high probability these selected items will have at
least one collision with the sorted items. If the quantum search
algorithm is run on these $O\left(N^{2/3}\right) $ items, it will
find the collision in $O\left(N^{1/3}\right) $ queries. Each query
takes only a logarithmic number of time steps since the
$O\left(N^{1/3}\right) $ items have been sorted. Thus the total
number of time steps required by the algorithm is
$O\left(N^{1/3}\right) $ steps to do the sorting plus
$O\left(N^{1/3}\right) $ steps to do the searching which is
$O\left(N^{1/3}\right) $ steps in all.

This algorithm achieves an $O\left(\frac{N^{1/2}}{N^{1/3}}\right)
=O\left(N^{1/6}\right) $ speedup over what a simple quantum search
would take, but at the cost of using $O\left(N^{1/3}\right) $
quantum hardware in the form of memory registers. This is exactly
the same speedup that would be obtained by using parallel
processors to run standard quantum searching.

\subsection{Element distinctness algorithms}

\subsubsection{Algorithm (i)}

A quantum algorithm that took only $O\left(N^{3/4}\right) $ time
steps is given in \cite{buhrman}. This used a two level quantum
search. At the top level, it divided the $N$ items into $\sqrt{N}$
groups of $\sqrt{N}$
items each and it ran a quantum search on the $\sqrt{N}$ groups which took $%
O(N^{1/4})$ queries. In each query, the algorithm sorts the
$\sqrt{N}$ items in the group and then runs a quantum search on
all $N$ items to check if any of the $\sqrt{N}$ items has the same
function value as any of the $N$ items. Since the $\sqrt{N}$ items
have been sorted, each check takes only a logarithmic number of
steps. Therefore each top level query takes $O\left(N^{1/2}\right)
$ steps to do the sorting plus $O\left(N^{1/2}\right) $ steps to
do the searching which is $O\left(N^{1/2}\right) $ steps in all.

This algorithm achieves an $O\left(\frac{N}{N^{3/4}}\right)
=O\left(N^{1/4}\right) $ speedup over what a simple quantum search
would take, but at the cost of using $O\left(N^{1/2}\right) $
quantum hardware in the form of memory registers. This is the same
speedup that would be obtained by using parallel processors to run
standard quantum searching.

\subsubsection{Algorithm (ii)}

Recently Ambainis has discovered an algorithm for element
distinctness which we believe takes $O\left(N^{2/3}\right) $ time
steps and requires $\ O\left(N^{2/3}\right) $ memory registers
\cite{vaz}. We do not know the details of this, however based on
these parameters this algorithm achieves an
$O\left(\frac{N}{N^{2/3}}\right) =O\left(N^{1/3}\right) $ speedup
over what a simple quantum search would take but at the cost of
using $O\left(N^{2/3}\right) $ quantum hardware in the form of
memory registers. Once again, this is exactly the same speedup
that would be obtained by using parallel processors to run quantum
searching as discussed above.
%

\subsection{Optimality of the algorithms?}

How fast could these algorithms possibly run? As indicated
previously, there are lower bounds known on how many time steps
are required. These are derived by lower bounding the number of
queries that an algorithm would need. This is because the number
of queries is generally considered the most convenient parameter
to use for analyzing the behavior of an algorithm. However, query
complexity is simply one way of characterizing the algorithmic
difficulty of a problem - the bottom line is the time it takes and
the hardware it uses. The algorithms above saturate the known
\emph{query complexity} lower bounds for the particular problems.
As we have seen, if we characterize these algorithm in terms of
more general space/time tradeoffs, they demonstrate no advantage
over parallel quantum searching.

\section{CONCLUSION}

We have argued that three well known algorithms for the collision
and element distinctness problems do not, in any meaningful way,
make algorithmic use of the problem structure to go beyond the
standard quantum searching paradigm. We therefore leave the reader
with the following:

\textbf{Challenge 1: }\emph{Find an algorithm for collision and/or element
distinctness which gives a searching speedup greater than merely a
square-root factor over the number of available processing qubits.}

\textbf{Challenge 2: }\emph{Find `physically significant' lower bounds for
collision and/or element distinctness - i.e. lower bounds in terms
of the total hardware/total time required as opposed to the less
meaningful bounds in terms of number of queries.}

\acknowledgements
This research was partly supported by NSA\ \&\ ARO under contract
no. DAAG55-98-C-0040.


\begin{thebibliography}{99}
\bibitem{grover96} L. K. Grover, \textit{Quantum Mechanics helps in searching for
a needle in a haystack}, \textit{Phys. Rev. Letters}, 78(2), \
325, 1997, also at http://www.bell-labs.com/user/lkgrover/.

\bibitem{bbbv} C. H. Bennett, E. Bernstein, G. Brassard \& U.Vazirani,
\textit{Strengths and weaknesses of quantum computing}, SIAM Journal on
Computing, 26, no. 5, Oct. 1997, p. 1510-1524.


\bibitem{zalka} C. Zalka, \textit{Grover's quantum searching is optimal,} Phys.
Rev. A 60, 2746 (1999).


\bibitem{amp} L. K. Grover, \textit{Quantum computers can search rapidly by
using almost any transformation}, Phys. Rev. Letters, 80(19),
1998, 4329-4332; \textit{A framework for fast quantum mechanical
algorithms}, Proc. 30th ACM Symposium on Theory of Computing
(STOC), 1998, 53-63.

\bibitem{BBHT} G. Brassard, P. Hoeyer, M. Mosca and Alain Tapp,
\textit{Quantum amplitude amplification and estimation},
quant-ph/0005055.

\bibitem{scott} S. Aaronson, \textit{Quantum lower bound for the collision problem},
STOC '02, pp.635-642. Also quant-ph/0111102.

\bibitem{shih} Y. Shih, \textit{Quantum lower bounds for the collision
and element distinctness problems}, quant-ph/0112086.


\bibitem{kutin} S. Kutin, \textit{Quantum lower bound for the collision
problem,}
quant-ph/0304162.

\bibitem{ambainis} A. Ambainis, \textit{Quantum Lower Bounds for Collision and
Element Distinctness with Small Range}, quant-ph/0305179.


\bibitem{BHT} G. Brassard, P. Hoyer \& A. Tapp, \textit{Quantum Algorithms for the
collision problem}, SIGACT News, 28:14-19,1997. Also
quant-ph/9705002

\bibitem{buhrman} H. Buhrman et al, \textit{Quantum Algorithms for Element
Distinctness}, quant-ph/0007016.


\bibitem{vaz} Umesh Vazirani, private communication.
\end{thebibliography}
\end{document}